\documentclass[float,preprintnumbers,amssymb,twocolumn,superscriptaddress,aps]{revtex4-2}
\usepackage{graphicx}
\usepackage{adjustbox}
\usepackage{booktabs}
\usepackage{amsfonts, amsmath, amsthm, amssymb,mathrsfs} 
\usepackage{mathtools}
\usepackage{wrapfig, tikz,url,lipsum,float}
\usepackage[normalem]{ulem}
\usepackage{bm}
\usepackage{subfigure}
\usepackage{color}
\usepackage[colorlinks=true,linkcolor=blue,citecolor=blue]{hyperref}
\usepackage{hyperref}
\usepackage{footnote}
\usepackage{cleveref}
\usepackage[utf8]{inputenc}
\usepackage{multirow}

\begin{document}



\title{\bf Mean first passage time of active fluctuating membrane with stochastic resetting}

\author{Tapas Singha} 
\email{tapas134@gmail.com}
\affiliation{Institut Curie, Université PSL, Sorbonne Université, CNRS UMR168, Laboratoire Physico Chimie Curie, 75005 Paris, France}

\begin{abstract}
We study the mean first passage time of a one-dimensional active fluctuating membrane that is stochastically returned to the same flat initial condition at a finite rate. We start with a Fokker-Planck equation to describe the evolution of the membrane coupled with an Ornstein-Uhlenbeck type of active noise. Using the method of characteristics, we  solve the equation and obtain the joint distribution of the membrane height and active noise. In order to obtain the mean first-passage time (MFPT), we further obtain a relation between the MFPT and a propagator that includes stochastic resetting. The derived relation is then used to calculate it analytically. Our studies show that the MFPT increases with a larger resetting rate and decreases with a smaller rate, i.e., there is an optimal resetting rate. We compare the results in terms of MFPT of the membrane with active and thermal noises for different membrane properties. The optimal resetting rate is much smaller with active noise compared to thermal. When the resetting rate is much lower than the optimal rate, we demonstrate how the MFPT scales with resetting rates, distance to the target, and the properties of the membranes.
\end{abstract}

\maketitle
\section{Introduction}
A biological membrane is comprised of a variety of proteins and ion channels that interact with the intra- and extra-cellular environments. The proteins and ion channels consume energy via the hydrolysis of adenosine triphosphate (ATP) and exert force on the membrane\cite{Manneville1999}. On average, the exerted active force could be either zero or nonzero, depending on the exact problem. The interplay of the force and the mechanical properties of the membrane, such as bending rigidity and tension, determines the shape of the membrane and the movement of a cell.
 
In order to perform a specific function, a cell often needs to reach a certain target. When it moves with an average velocity, calculating the time it will take to reach a target seems straightforward. We consider a scenario in which a cell finds its static single target only through membrane fluctuations. In this case, it is interesting to understand the time a membrane takes to reach its target for the first time. As the membrane fluctuates either towards or away from the target, the reaching time is a stochastic variable, and in a statistical sense, the mean first-passage time may be a sensible measure for the process. 

The mean first-passage time (MFPT) is relevant in a variety of physical, chemical, and biological processes ranging from species extinctions in ecology\cite{Hanggi1990, Redner2001} to molecular processes. A simple example of MFPT is the average time a forager takes to find food or other resources for the first time\cite{Bell1991}. In biology, the immune cells search for cells with antigens. Therefore, it is relevant in a range of length scales and becomes more important as the need to understand complex systems increases. 

Because the underlying stochastic dynamics of a fluctuating system can cause MFPT to be infinitely long, depending on the system details, several search strategies have been proposed over the last two decades\cite{Oshanin2009, Bartumeus2002,Viswanathan1996}. One of the most important strategies is stochastic resetting, in which the searcher is returned to its initial state at a fixed rate.  This may enhance the likelihood of finding the target because the searcher may explore the surroundings, that may reduce the likelihood of wandering away from the target. Stochastic resetting was first proposed for a single Brownian particle\cite{Evans2011, Evans2011_JPA,Evans2013}, subsequently, numerous problems have also been studied in other fields, such as population expansion in fluctuating environments\cite{Kussell2005} and  search algorithms in computer science\cite{Montanari2002}. Several variations of resetting\cite{Nagar2016} have also been investigated, including resetting with finite velocity\cite{Tal-Friedman2020}, which may be useful in real life. 
 
 \begin{figure}[H]
\hspace{0.04cm}
        \includegraphics[width=0.85\linewidth, height=0.16\textheight]{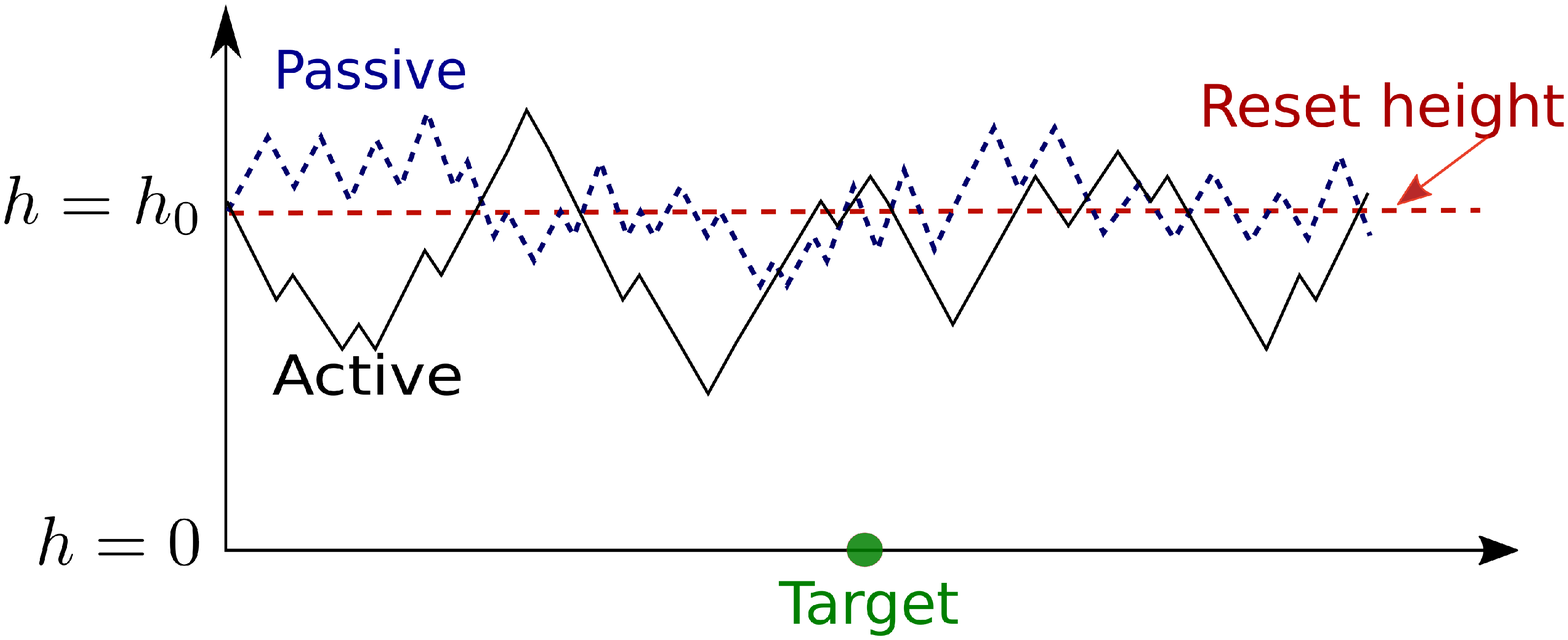}
\caption{Schematic diagram of fluctuating active (curvy, solid black line) and passive (curvy, dashed blue line) membranes and their target (shown by a solid green circle).The membrane is reset to its initial height $h=h_0$ (shown by a straight-dashed red line) at a finite rate.}\label{fig:ResetMembrane}
\end{figure}

A passive membrane driven by thermal noise satisfies the fluctuation-dissipation theorem. A biological membrane, on the other hand, transports ions through its embedded ion pumps, which consume ATP and cause local force (or active noise) on the membrane. Together with thermal noise, this active noise enhances the amplitude of membrane fluctuations, which can be experimentally quantified by the effective temperature\cite{Manneville1999,Manneville2001}. Unlike the passive membrane, the membrane, which actively participates in biological processes, does not satisfy the fluctuation-dissipation theorem. 

Though the passive surface growth with stochastic resetting  has been studied\cite{Gupta2014}, the dynamics of a membrane with active fluctuations have yet to be investigated. Importantly, the MFPT, which is fascinating in the context of searching, remains elusive. In this work, we answer the question: how long does it take for a membrane to reach its single point target when a membrane is reset to its flat initial height with a finite rate? We illustrate the scenario in the schematic diagram shown in Fig.\,(\ref{fig:ResetMembrane}). 

To this end, we first derive the propagator for the coupled dynamical equations for an active membrane. Further, we derive a general relation between the MFPT and a propagator under resetting. Using the relation and the derived propagator for the height, we obtain the MFPT. Our studies reveal that there exists an optimal resetting rate at which MFPT becomes minimum.  We demonstrate how, for different membrane properties, MFPT scales with resetting rates and target heights, and how an optimal resetting rate arises. Finally, we compare the MFPT for active and passive membranes.

The resetting of a membrane may be related to the type of motility in which a part of the cell membrane gets detached from its cortex and fluctuates, known as a bleb. After that, the cortical actin filaments reach the detached membrane and bring the membrane back to its initial position\cite{Charras2008,Brugues2010}. The membrane may grow again and find its target; however, the underlying dynamics remain unclear.

This work is structured as follows: The model is presented in Section II. The derivation of the single-height distribution is described in Section III. Section IV depicts the connection between the mean first-passage time and the propagator. In the following section, we calculate the height-height correlations for different cases. Section VI presents the MFPT for a variety of cases in one spatial dimension. Finally, Section VII presents discussion and conclusions.

\section{Model}
We are interested in understanding how the symmetric membrane fluctuations eventually drive the membrane to reach a target. We consider a membrane with the symmetric distribution of all membrane components (such as lipids and proteins) and the same environment on both sides of the membrane that may ensure symmetric membrane fluctuations and, thereby zero spontaneous curvature. Here we first consider the free energy for a symmetric membrane \cite{Helfrich1973} expressed as 
\begin{align}
\mathcal{F}[h] = \int d x \left \{ \frac{\nu}{2} (\nabla h)^2 + \frac{\kappa}{2} (\nabla^2 h)^2  \right\} \label{HelfrichPot}    
\end{align}
where $\nu$, and $\kappa$ are the tension and the bending rigidity of the membrane, respectively. We study a simple dynamical equation for height field $h(x,t)$ in one spatial dimension written as  $\Gamma \partial  h/\partial t = - \delta \mathcal{F}[h]/\delta h  +  \eta (x,t) + a v(x,t), $
where $\eta(x,t)$ is the thermal noise with zero average and $\langle \eta(x,t) \eta(x',t') \rangle= 2 D \delta (x-x') \delta(t-t').$ The coefficient $D$ is the strength of the thermal noise. From here onwards, we scale time $t$ with $\Gamma$, and keep  it as $t$. The last term  $v(x, t)$ arises from the fluctuating force  of ion pumps/channels and the proteins which affect the membrane dynamics. Considering $v(x,t)$ as an Ornstein-Uhlenbeck type of field, we write
\begin{equation}
\frac{\partial v (x,t)}{ \partial t} = - \frac{v}{\tau_{a}} + \mu (x,t)
\label{vel_noise}
\end{equation}
where $\mu(x,t)$ is the Gaussian noise
with zero average and $\langle \mu(x,t) \mu(x',t') \rangle= 2 D_a \delta (x-x') \delta(t-t')$, and $\tau_a$ is the relaxation time of an active noise field $v$. The strength of the active fluctuations $D_{a}$ depends on the available ATP concentration, and the density of the protein pumps \cite{Gov2004} which are taken to be constant in space and time. As there are no spatial derivatives or other functions of space in   Eq.\,(\ref{vel_noise}), $v_i$ acts locally on the membrane $h_i$. In contrast to thermal noise, active noise $v_i$ has a temporal correlation with a correlation time $\tau_a$ resulting in membrane height $h_i$ being kicked for a much longer period of time than that of thermal noise. To put it another way, the active noise correlation effectively takes the form of the thermal noise correlation in the limit $\tau_a \rightarrow 0$.

When a membrane is surrounded by fluid, the motion at one point of the membrane affects the motion at the other point via the fluid medium, which is known as hydrodynamic interaction. However, in one spatial dimension, the effect of the hydrodynamic interactions on the relaxation dynamics is marginal\cite{Farge1997,Granek1997}. We, therefore, neglect the hydrodynamic interactions and write the dynamical equation for membrane height as 
\begin{equation}
\frac{\partial h(x,t)}{\partial t} =  \nu \,\frac{\partial^2  h}{\partial x^2} - \kappa\,\frac{\partial^4  h}{\partial x^4}   +a\, v(x,t) + \eta (x,t). \label{AEW}
\end{equation}

In the above equation, bending rigidity $\kappa$ and tension $\nu$ determine a length scale $\ell_c=\sqrt{\kappa/\nu}$. As a result, when $\ell> \ell_c$, $\nu$ dominates membrane dynamics; otherwise, $\kappa$ dominates. We study the two limits of the membrane properties, namely tension-less active membrane (TLAM) ($\nu = 0$) in which dynamics are governed by bending rigidity $\kappa$, and tension-dominated active membrane (TDAM) ($\kappa=0$) in which dynamics are governed by tension $\nu$. To study active dynamics, we neglect the thermal noise $\eta$ compared with active noise $v$, and we study the membrane dynamics separately with thermal noise.  

The growth of the height fluctuations may vary over time. For instance, the initial height fluctuation dynamics differ from the late-time dynamics. The mean-squared width, which is a statistical measure of height fluctuations, grows with $t^{\beta}$ for $t \ll L^z$, where $\beta$ and  $z$ are the growth and the dynamic exponents, respectively, and $L$ represents the size of the system. For $t\gg L^{z}$, the system reaches its steady state, and mean-squared width no longer depends on time $t$, rather on $L^{\chi}$ where $\chi$ is the roughness exponent of the membrane \cite{BarabasiStanley1995}. The dynamic exponent $z$ is determined by the properties of the membrane; for example, when tension dominates membrane relaxation ($\nu \neq 0$ and $\kappa=0$)  the dynamic exponent $z=2$ and when bending rigidity dominates membrane relaxation ($\kappa \neq 0$ and $\nu = 0$), $z=4$. For the active system, the timescale of interest is expressed as $t \ll \tau_a \ll L^{z}$ for $L \rightarrow \infty.$



\section{Height distribution}
We first aim to derive the propagator for the coupled equations given in Eqs.\,(\ref{vel_noise}), and (\ref{AEW}). The discretized versions of Eqs.\, (\ref{vel_noise}) and (\ref{AEW}) are written as 
\begin{equation}
\frac{\partial }{\partial t} \begin{bmatrix}
h_i(t) \\
v_i(t)
\end{bmatrix} 
= \begin{bmatrix} 
              -\sum_{j}\Lambda_{ij}  & a\\ 0 & -\Lambda'
             \end{bmatrix}
             \begin{bmatrix}
             h_j(t) \\
             v_i(t)
            \end{bmatrix} + \begin{bmatrix}
\eta_i(t) \\
\mu_i(t)
\end{bmatrix}
\end{equation} 
where $\Lambda_{ij} = -(\nu \Delta_{ij}-\kappa \Delta_{ij}^2)$, $\Lambda' = 1/\tau_a$, and $\sum_{j}$ in the above equation includes the nearest neighbors. Further, we write the Fokker-Planck equation for the joint distribution $\mathbf{W} (\{h\},\{v\},t|h^0,v^0,t_0)$ as 
\begin{eqnarray}
&& \frac{\partial \mathbf{W}}{\partial t} = [-\sum^{N}_{i,j=1} \frac{\partial}{\partial h_{i}}(- \Lambda_{ij} h_j + a\, \delta_{ij}\,v_j) +\sum^{N}_{i,j=1} \frac{\partial}{\partial v_i} (\Lambda_{ij}'v_j) \nonumber \\ && + \sum^{N}_{i,j=1} \frac{\partial^2}{\partial h_i \partial h_j} D_{ij} +  \sum^{N}_{i,j=1} \frac{\partial^2}{\partial v_i \partial v_j} D^a_{ij} ] \,\,  \mathbf{W} (\tilde{h},\tilde{v},t)
\end{eqnarray}
where $N$ is the total number of sites. The initial conditions read $ \mathbf{W}(\tilde{h},\tilde{v},t_0|h^{0},v^{0},t_0) = \delta(\tilde{h}-\tilde{h}^{0}) \,\, \delta(\tilde{v}-\tilde{v}^{0})$ where  $\tilde{h} \equiv \{ h_i\}$ and $\tilde{v} \equiv \{v_i\}$.  Using the method of characteristics, we solve the above equation (as detailed in \emph{Appendix-A}). Further, we integrate the distribution of initial noise field $\tilde{v}^{0}$ as $\int d \tilde{v}^{0} \, \delta(\tilde{v}^{0})\,  \mathbf{W}(\tilde{h},t|\tilde{\psi}^{0}) $. We also integrate out the active noise field $\tilde{v}$ and then obtain a marginal distribution for $\tilde{h}$ which reads as
\begin{eqnarray}
W(\tilde{h},t|0,0) && = \frac{\exp{[-\frac{1}{2}\tilde{h}^T \mathbf{M}^{-1}_1 \tilde{h}]}}{(2\pi)^{L/2} \sqrt{\det{\mathbf{M}}_1}} 
\label{ProbDist_Ht}
\end{eqnarray}
where covariance matrix $\mathbf{M}_1 $ is expressed as
\begin{eqnarray}
\mathbf{M}_1= \left[2D_{\text{tot}}f(2\Lambda) + a'^2(f(2\Lambda')-2f(\Lambda+\Lambda'))\right] \delta_{ij} 
\label{M1Matrix}
\end{eqnarray}
in which $f(\theta)=(1-e^{-\theta\,t})/\theta$, $D_{\text{tot}}=D+a'^2/2$, and $a'^2=2a^2D_a/(\Lambda'-\Lambda)^2$. The above matrix is obtained with the non-stationary active noise. Considering $a=0$ in Eq.\,(\ref{ProbDist_Ht}), we obtain the propagator for the passive system  expressed as $W_{\text{passive}}(\tilde{h},t|0,0) = \sqrt{\frac{1}{(2\pi)^{L}}\det{\left(\frac{\Lambda}{D\,(1-e^{-2\Lambda t})}\right)}}\,\exp{\left[-\frac{1}{2}\, \tilde{h}^T \frac{\Lambda}{D(1-e^{-2\Lambda t})} \tilde{h}\,\right]},
$ which is consistent with the propagator for the Ornstein-Uhlenbeck type of particle in a harmonic potential \cite{RiskenBook}.

As we define the position of the target as being at $0$,  we integrate out heights from all the sites except $x=0$ because a single height at that point may only reach the target as overhangs are not considered. For a homogeneous system, we obtain the marginal distribution  
\begin{eqnarray}
W(\mathbf{0},t|h_0,0) &&= \frac{1}{\sqrt{2 \pi \langle h^2(t) \rangle}}  \exp{\left[-\frac{1}{2} \frac{h_0^2}{ \langle h^2(t) \rangle}\right]}.  \nonumber \\
\label{EqPropagator}
\end{eqnarray}  
The obtained propagator in the above equation describes how a single height distribution evolves with time starting from a reference height $h_0$.

\section{Mean First Passage time with resetting}
We next present a relation between the first-passage time and survival probability, and then investigate how the mean first-passage time is related to a propagator.

\subsection{Survival probability}
Let us consider a  particle driven by stochastic noise starts at position $h=h_0$, and reaches the target ($h=0$) for the first time at time $T$, known as first-passage time. The survival probability, $S(h_0, T)$, can be related to the first-passage time distribution, $F(h_0, T)$, which is defined as the likelihood that the searcher has not reached the target up to time $T$. With this, $F(h_0, T)$ can be expressed by the difference between the survival probabilities at two consecutive discrete times as $F(h_0,T)= S(h_0,T-1)-S(h_0,T)$
\cite{Majumdar2010}. In the continuum limit $\Delta T \rightarrow 0$, $F(h_0,T) \Delta T = S(h_0,T-\Delta T)- S(h_0,T)$ yields 
\begin{equation}
 F(h_0,T) = - \frac{\partial S  (h_0,T)}{\partial T}. 
\end{equation}
We next study the MFPT of a stochastic variable in one spatial dimension under stochastic resetting.

\subsection{MFPT with Resetting}
 We consider that the initial separation between a point on the membrane and a target is $h_0$, and that the membrane takes time $T$ to reach the target for the first time (see fig.1).  Resetting is the immediate return of the membrane to its starting flat initial height $h_0$ at a rate $r$. The survival probability with resetting is denoted as $S_r(h_0, T)$ and the corresponding first-passage distribution as $F_r(h_0, T)$.  The first-passage distribution and the survival probability with resetting are related in the following way \cite{Gupta2022}: 
\begin{equation}
F_{r}(h_0,T) =  -\frac{\partial S_r(h_0, T)}{\partial T}.   
\label{RelFPT_SPD}
\end{equation}
Therefore, the mean first-passage time with resetting is defined as $
\langle T \rangle = \int^{\infty}_{0}  dT\, T F_r(h_0,T). $ Using Eq.\,(\ref{RelFPT_SPD}) in the definition of the MFPT,  we get
\begin{equation}
\langle T \rangle = \int^{\infty}_{0}  dT\,  S_r(h_0, T).
\label{MFPT_SP}
\end{equation}
Now, we consider the Laplace transform (LT) of survival probability, which reads as 
\begin{equation}
\widetilde{S}_r(h_0, s) = \int_0^{\infty} dT \, e^{-s \, T}  \, S_r(h_0, T).
\label{LT_SurvProb}
\end{equation}
When $T \rightarrow \infty$, we consider that the searcher finds the target with probability $1$, which results in $S_r(h_0,\infty)=0$. Substituting $s=0$ in Eq.\,(\ref{LT_SurvProb}) and equating with Eq.\,(\ref{MFPT_SP}), we obtain 
\begin{equation}
\langle T \rangle = \widetilde{S}_r(h_0, 0),
\label{MFPT_LT_SurvivalProb}
\end{equation}
where $S_r(h_0,\infty)=0$ is considered. Following the backward Kolmogorov equation for first-passage time distribution, we write
\begin{eqnarray}
F_r(h_0, T) = e^{-r T} F(h_0, T) +  \int_0^{T} && dt'  S_r(h_0,  T-t')  \, r\nonumber\\ && \times\, e^{-r t'} F(h_0,t'),
\end{eqnarray}
where $F_r(h_0, T)$ is the first-passage time distribution with resetting. The first term on the right-hand side of the above equation indicates that there has been no resetting in time $T$, while the second term denotes that the last resetting took place at a time $t'$.

Using Eq.\,(\ref{RelFPT_SPD}), we multiply $e^{-s T}$ on both sides of the above equation and finally integrate over $T$. This leads to \begin{equation}
\widetilde{S}_r(h_0, s) = \frac{1-\widetilde{F}(h_0,r+s)}{s+r \widetilde{F}(h_0,r+s)}
\label{RelSP_FPT}
\end{equation}
where $\widetilde{F}(h_0,r+s)$ is the LT of the first-passage distribution time \emph{without} resetting. In the above equation, we use  $S_r(h_0,0)=1$, i.e., initially ($t=0$) the survival probability is $1$. Substituting $s=0$ in the above equation, and using Eq.\,(\ref{MFPT_LT_SurvivalProb}), we obtain
\begin{eqnarray}
\langle T \rangle = \frac{1}{r} \left( \frac{1}{ \widetilde{F}(h_0,r)} -1\right). 
\label{MFPT_FPD}
\end{eqnarray}
Following the renewal equation, we write the relation between the first-passage time distribution and the propagator as \cite{Redner2001, Hughes1995} 
\begin{eqnarray}
W(\mathbf{0},T|h_0) = \int^{T}_0 dt' F(h_0,t') W &&(\mathbf{0},T-t'|\mathbf{0}) dt'  \nonumber \\
&& + \delta_{(0,h_0)} \delta(T)
\end{eqnarray}
where $W(\mathbf{0},T|h_0)$ is the propagator at  $h=0$ at time $T$ given that the initial coordinate was at $h=h_0$. The Laplace transform of the above equation leads to
\begin{equation}
\widetilde{F}(h_0,s) = \frac{\widetilde{W}(\mathbf{0},s|h_0)}{\widetilde{W}(\mathbf{0},s|\mathbf{0})}
\end{equation}
where $h_0 \neq 0$, and $s$ is the Laplace conjugate to time $T$. Substituting the above equation in 
Eq.\,(\ref{MFPT_FPD}), we obtain
\begin{eqnarray}
\langle T \rangle = \frac{1}{r} \left( \frac{\widetilde{W}(\mathbf{0},r|\mathbf{0})}{\widetilde{W}(\mathbf{0},r|h_0)}  -1\right) .
\label{MFPT_LT_Prop}
\end{eqnarray}

Despite the fact that we discuss the case of a point on the membrane, the obtained relation is valid for any stochastic variable. 

\section{Height-height correlation }
The auto-correlation of active proteins from Eq.(\ref{vel_noise}) can be written as   
 \begin{eqnarray}
\langle v_{i}(t)v_{j}(t') \rangle &&= e^{-\Lambda'(t+t')} \int^{t}_{t_0} ds \int^{t'}_{t_0} ds' e^{\Lambda'(s+s')} \langle \mu_{i}(s) \mu_{j}(s') \rangle \nonumber \\
\end{eqnarray}
where $\langle v_i(t) \rangle =0$, and $t_0$ is the initial time. We consider that the active proteins have already reached their steady state as they relax faster than the membrane, which is achieved by setting $t_0 \rightarrow -\infty$. Carrying out the integration in the above equation, we obtain
\begin{eqnarray}
\langle v_{i}(t)v_{j}(t') \rangle
 = \, D_a \,  \frac{e^{-\Lambda' |t-t'|}}{\Lambda'}  \delta_{ij}.
 \label{station_ActiveNoise}
\end{eqnarray}
The derivation of  $\mathbf{M}_1$ (given in Eq.\,(\ref{M1Matrix})) from the Fokker-Planck equation and $\langle h_{i}(t) h_{j}(t) \rangle$ from the coupled Langevin equations (given in Eqs.\,(\ref{vel_noise}),(\ref{AEW})) are found to be the same\begin{equation}
\langle h_{i}(t) h_{j}(t) \rangle = M^{ij}_{1}\,\,  \delta_{ij}.
\end{equation}
The above relation is obtained with the non-stationary and stationary state active noise. With non-stationary state active noise, the relation is shown in Appendix-A, and with stationary state, the height correlation is obtained as
\begin{eqnarray}
\langle h^2_i(t)\rangle =2 \left(D + D'_a\right) f(2\Lambda) && -  2 D'_a \, f(\Lambda+\Lambda'), \label{ht_corr_SS}
\end{eqnarray}
where $D'_a = \frac{a^2D_a}{\Lambda'(\Lambda'-\Lambda)}.$ When $\Lambda$ is considered as the spring constant of a harmonic well, the above equation is consistent with the displacement correlation of a particle in a harmonic well with an Ornstein-Uhlenbeck type of active noise \cite{Das2018}. For $\Lambda \rightarrow 0$, Eq.\,(\ref{ht_corr_SS}) becomes equivalent to the displacement correlation of the active Ornstein-Uhlenbeck type of noise. Further, considering $\Lambda'=1/\tau_a$, the above correlation can be obtained as follows: $ \lim_{\Lambda \rightarrow 0}\langle h^2_i(t)\rangle=2t\left(D+(a\tau_a)^2 D_a \left[1-\frac{\tau_a}{t}(1-e^{-t/\tau_a}) \right]\right)$ which agrees with ref.\cite{Goswami2019}. In the passive case ($a=0$), the above correlation becomes $2Dt$, which is the mean-square displacement of a Brownian particle in one dimension.

In order to investigate the MFPT of fluctuating membranes, we first employ the relation between the MFPT and the propagator given in Eq.\,(\ref{MFPT_LT_Prop}). We next aim to obtain the propagator with active or thermal noise (as given in Eq.\,(\ref{EqPropagator})), which requires the single point height-height correlation. To this end, we begin with the Fourier transform of Eq.(\ref{AEW}) which reads as \begin{eqnarray}
\frac{\partial h(q,t) }{\partial t} = - \frac{1}{\tau_q} h(q, t) +\eta(q, t) +a\, v(q, t)
\label{FT_AEW}
\end{eqnarray}
where $\,\tau_q = (\nu q^2+\kappa q^4)^{-1}$ is the membrane relaxation. Using the flat initial condition, we get 
\begin{eqnarray}
h(q,t) = e^{-t/\tau_q} \int^{t}_{0} dt' \, e^{t'/\tau_q} \,\, [\eta (q,t') +a\, v(q,t')].
\label{Active_Vel}
\end{eqnarray}
We also obtain the auto-correlation of the stationary state active noises as 
\begin{equation}
\langle v(q,t)v(q',t')\rangle = D_a \tau_a\,  (2\pi) \delta(q+q')  \,  e^{-|t'-t|/\tau_a}.   
\label{SS_ActiveNOiseCorre}
\end{equation} 
The inverse Fourier transform of the height-height correlation can be written as 
\begin{eqnarray}
\langle h(x,t)^2 \rangle
&& = \int^{\infty}_{-\infty} \frac{dq}{(2\pi)} \int^{\infty}_{-\infty} \frac{dq'}{(2\pi)} \,\, e^{i (q+q') x}\, \langle h(q,t) h(q',t) \rangle  \nonumber \\
\label{HtFlucActive}
\end{eqnarray} in which the correlation can be calculated either with active or thermal noises. 

\subsection{Tension-dominated membrane ($\kappa=0$)}
\subsubsection{Active}
Using the active noise correlation given in Eq.\,(\ref{SS_ActiveNOiseCorre}) and considering the Fourier transform of the height given in Eq.\,(\ref{Active_Vel}), we write the height-height correlation as
\begin{eqnarray}
\langle h(x,t)^2 \rangle_{\text{active}} = a^2\,D_a \, \tau_a  \int^{\infty}_{-\infty} && \frac{d q}{(2\pi)}  e^{-2\nu t q^2}  \int^{t}_{0}du  \int^{t}_{0} du'  \nonumber \\ && \times e^{\nu(u+u')q^2} e^{-|u-u'|/\tau_a}.
\label{TD_Inte}
\end{eqnarray} 
We perform  integration over $q$ and obtain
\begin{eqnarray}
\langle h(x,t)^2 \rangle_{\text{active}} &&= \frac{a^2\,D_a \, \tau_a}{2\sqrt{\pi \nu}}\, \int^{t}_{0} du \int^{t}_{0} du' \, \frac{e^{-|u-u'|/\tau_a}}{\sqrt{2t-(u+u')}}.  \nonumber \\
\end{eqnarray}
Next, we carry out the integration over $u$ and $u'$ which results in
\begin{eqnarray}
\langle h(x,t)^2 \rangle_{\text{active}}  
&&=\frac{a^2D_a \tau^{3/2}_a}{\sqrt{\nu}} ( \,[\sqrt{\frac{2 t \tau_a }{\pi }}-\frac{\tau_a}{2}  \, \text{Erf}\left(\sqrt{\frac{t}{\tau_a }}\right)] \nonumber \\ + && \frac{\tau_a}{2}   e^{-\frac{2 t}{\tau_a}} [\text{Erfi}\left(\sqrt{\frac{t}{\tau_a }}\right)-\text{Erfi}\left(\sqrt{\frac{2 t}{\tau_a }}\right)]). \nonumber \\
\label{TDAM_Ht-HT}
\end{eqnarray}
In this study, we are interested in the regime in which active noise drives the membrane. As we see above, depending on time $t$ and $\tau_a$, the active noise correlation given in Eq.\,(\ref{TDAM_Ht-HT})  has two distinct regimes. For $t\ll \tau_a$, the above equation is simplified to 
\begin{eqnarray}
    \langle h(x,t)^2 \rangle_{\text{active}} \simeq \frac{4}{3} (\sqrt{2}-1)\,\frac{a^2\,D_a \tau_a}{\sqrt{\pi \nu}}\, t^{3/2} \label{ActiveHt-HtCorr} 
\end{eqnarray}
and for $t\gg \tau_a$
\begin{eqnarray}
 \langle h(x,t)^2 \rangle_{\text{active}} \simeq a^2\,D_a \tau^2_a\, \sqrt{\frac{2t}{\pi \nu}}.
\end{eqnarray}

\subsubsection{Passive}
Similarly, we next derive  the auto-correlation with thermal noise (see Appendix B) which reads as
\begin{eqnarray}
\langle h(x,t)^2 \rangle_{\text{passive}}  = D \sqrt{\frac{2t}{\pi \nu}}.
\label{HtHt_ThermalNu}
\end{eqnarray}
In the limit $t \gg \tau_a$, the auto-correlation of active noise exhibits passive-like behavior. Because the active noise becomes uncorrelated when the interval between two successive observations is larger than the correlation time $\tau_a$. As a result, the effect of several changes in the uncorrelated noise $v_i$ changes the membrane height $h_i$ in a random manner. Thus, in the long time limit, this uncorrelated active noise becomes additive to the thermal noises, and according to the central limit theorem, the distribution of these uncorrelated noises leads to a Gaussian distribution with increased variance.

Since we study the passive case separately, for active noise, we limit ourselves only when $t \ll \tau_a$. 

\subsection{Tension-less membrane ($\nu=0$)}
\subsubsection{Active}
\begin{figure}[H]
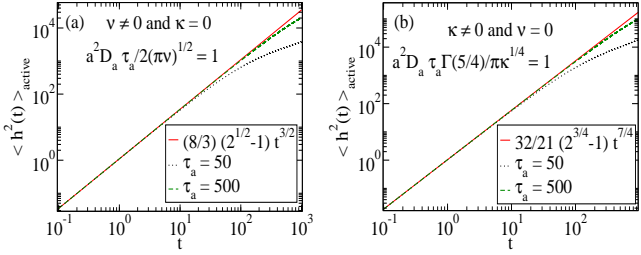

\centering
\begin{minipage}{0.5\textwidth}
\centering
        \includegraphics[width=0.46\linewidth, height=0.14\textheight]{EW.eps}%
        \hspace{0.1cm}
        \includegraphics[width=0.46\linewidth, height=0.14\textheight]{MH.eps}
\end{minipage}%
\caption{\,(a) The integration given in Eq.(\ref{TD_Inte}) for tension-dominated membrane fluctuations is numerically solved for $\tau_a=50 \,\, \text{and}\,\, 500$ (shown by the dotted-black and dashed-green lines, respectively), and the results are compared to the analytical findings given in Eq.\,(\ref{ActiveHt-HtCorr}) (shown by the solid-red line).  (b) Similarly, the integration given in Eq.(\ref{BD_Inte}) for rigidity-dominated membrane fluctuations is numerically solved (shown by the dotted-black and dashed-green lines, respectively), and the results are then compared with Eq.\,(\ref{ActiveCorrKappa}) (shown by the dashed-red line).}
\label{Fig_Active}
\end{figure}

When the membrane tension is very low compared to the bending rigidity, the dynamics of membrane relaxation are dominated by $\kappa$. We investigate how the height-height correlations scale in time for tension-less active membrane (TLAM). We begin with the height-height correlation expressed as 
\begin{eqnarray}
\langle h(x,t)^2 \rangle_{\text{active}} = a^2\,D_a \, \tau_a \int^{\infty}_{-\infty}  \frac{d q}{(2\pi)}  && \int^{t}_{0}du  \int^{t}_{0} du' e^{\kappa(u+u'-2t)q^4} \nonumber \\ && \times \, e^{-|u-u'|/\tau_a} .
\label{BD_Inte}
\end{eqnarray}
Integrating over $q$, we obtain 
\begin{eqnarray}
\langle h(x,t)^2 \rangle_{\text{active}} = \Gamma\left(\frac{5}{4}\right) \frac{a^2D_a \, \tau_a }{\pi \kappa^{1/4}}\, \int^{t}_{0} du \int^{t}_{0} du' \, \frac{e^{-|u-u'|/\tau_a}}{[2t-(u+u')]^{\frac{1}{4}}} .\nonumber \\
\end{eqnarray}
Next, integrating over $u$ and $u'$ and considering $t \ll \tau_a$, we obtain 
\begin{eqnarray}
\langle h(x,t)^2 \rangle_{\text{active}} && \,\simeq (2^{3/4}-1)\frac{8\,\Gamma(\frac{1}{4})}{21}  \frac{a^2\,D_a \, \tau_a }{\pi \kappa^{1/4}}\,\, t^{7/4} .
\label{ActiveCorrKappa}
\end{eqnarray}
We now carry out numerical integrations given in Eqs. (\ref{TD_Inte}) and (\ref{BD_Inte}), and compare the numerical results with the obtained analytical results  given in Eqs.\,(\ref{ActiveHt-HtCorr}) and (\ref{ActiveCorrKappa}), respectively for $t \ll \tau_a$ (shown in Fig.\,(\ref{Fig_Active}))

\subsubsection{Passive}
We also obtain the height-height correlation for the passive membrane as
\begin{eqnarray}
\langle h(x,t)^2 \rangle_{\text{passive}} \, &&= \frac{D}{3\pi^{3/4}}\, \left[ \frac{(2t)^3}{\pi \kappa} \right]^{1/4}.
\label{PassiveCorrKappa}
\end{eqnarray}

 With active noise, the height-height correlation  ($\sim t^{7/4}$) grows much faster than the thermal noise ($\sim t^{3/4}$), on the other hand, the scaling of $\kappa$ remains the same ($\sim \kappa^{-1/4}$) for both the noises. We obtain similar results for tension-dominant membranes in which the correlation $\langle h^2 \rangle \sim \nu^{-1/2}$ for both the noises, whereas the time dependence of the correlation gets significantly affected by the active noise as given in Eqs.\,(\ref{ActiveHt-HtCorr}) and (\ref{HtHt_ThermalNu}).


\section{Mean First-Passage Time}
In this section, we derive the MFPT for TDAM via the LT of the propagator. 

\subsection{\bf MFPT: Tension-dominated active membrane}  As the diagonal matrix $\mathbf{M}_1$ is the exactly same as $\langle h^2_i(t)\rangle$, we substitute Eq. (\ref{ActiveHt-HtCorr}) in Eq.\,(\ref{EqPropagator}), and obtain 
\begin{equation}
W_{\text{active}}(h_0,t) = \frac{1}{\sqrt{\pi}}  \frac{1}{\alpha_t \,}   \exp{\left(-\frac{h_0^2}{\alpha_t^2}\right)} 
\label{AEW_Propa}
\end{equation}
where $\alpha_t^{-1}=\sqrt{3 \sqrt{\pi \nu}\,}/\left(\sqrt{8  a^2D_a \tau_a (\sqrt{2}-1)} \,\, t^{3/4}\right).$ 
The Laplace transform of Eq.\,(\ref{AEW_Propa}) yields  
\begin{equation}
\widetilde{W}_{\text{active}}(u_a,r) = \frac{1}{\sqrt{\pi} \alpha} \frac{1}{r^{1/4}}\,\int^{\infty}_{0} \, \frac{d\bar{t}}{\bar{t}^{3/4}}  \exp{\left(-\bar{t}-\frac{u_a^2}{\bar{t}^{\frac{3}{2}}} \right)}
\end{equation}
where $\alpha = \alpha_t/t^{3/4}$, $\bar{t}=r\,t$ and $u_a= r^{3/4}\, h_0/\alpha$. We substitute the above equation in Eq.\,(\ref{MFPT_LT_Prop}), and obtain 
\begin{equation}
\langle T_{\nu} \rangle_{\text{active}} = \frac{1}{r}  \left(\frac{I_a(0)}{I_a(u_a)} -1 \right)
\label{Time_TDAM}
\end{equation}

\begin{figure}
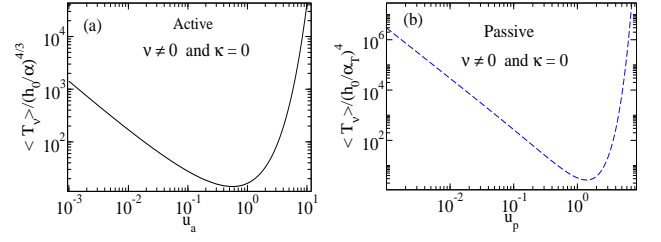

\begin{minipage}{.5\textwidth}
\centering
        \includegraphics[width=0.45\linewidth, height=0.13\textheight]{MFPTActiveNu.eps}%
        \hspace{0.1cm}
        \includegraphics[width=0.45\linewidth, height=0.13\textheight]{MFPTPassiveNu.eps}
\end{minipage}%
\caption{The variation of mean first-passage time $\langle T \rangle$ with resetting rate $r$ for a tension-dominated membrane ($\nu \neq 0$ and $\kappa = 0$) and fixed target height $h_0$. (a) For an active membrane, we plot the MFPT, which is analytically obtained in Eq.(\ref{Time_TDAM}), where $u_a=r^{3/4} h_0\,\sqrt{(3\sqrt{\pi \nu})/(8 a^2\,D_a \tau_a (\sqrt{2}-1))}.$ (b) For passive membrane, we plot MFPT obtained in Eq.(\ref{Time_TDPM}) where $u_p=r^{1/4} h_0/(2 D \sqrt{2/(\pi \nu)})^{1/2}$  as the effective coupling variables for active (shown in left figure (a)) and passive (shown in right figure (b)) systems, respectively. In contrast to the passive membrane, where the effective coupling rate $u_p>1$, the active membrane has an optimal resetting rate for $u_a<1$.}
\label{figTensionDominated}
\end{figure}

\begin{figure}
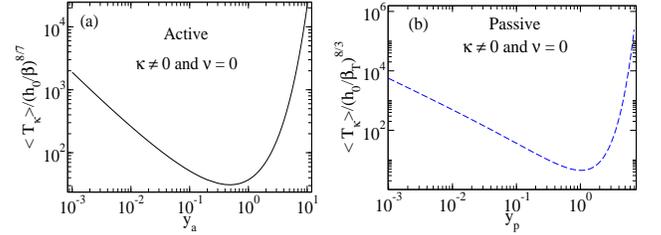

\begin{minipage}{.5\textwidth}
\centering
        \includegraphics[width=0.45\linewidth, height=0.13\textheight]{MFPTActiveKappa.eps}%
         \hspace{0.1cm}
        \includegraphics[width=0.45\linewidth, height=0.13\textheight]{MFPTPassiveKappa.eps}
\end{minipage}%
\caption{Variation of mean first-passage time $\langle T \rangle$ with resetting rate $r$ for a tension-less membrane ($\kappa \neq 0$ and $\nu = 0$) and fixed target height $h_0$. (a) For active membrane, we  plot the MFPT via numerical integration which is analytically obtained in Eq.(\ref{Time_Tension-LessActive}) where $y_a= r^{7/8} h_0/\sqrt{(16/21) (2^{3/4}-1) \Gamma(1/4)a^2\,D_a \tau_a/\pi \kappa^{1/4}}$. (b) For passive membrane, we plot MFPT obtained in Eq.(\ref{Time_TLPM}) where $y_p=r^{3/8} h_0 \sqrt{3\pi \kappa^{1/4}/(2^{7/4} D)}$ as the effective coupling variables for active (shown in left figure (a)) and passive (shown in right figure (b)) systems, respectively. The optimal resetting rate is obtained for $y_a <1$ for active system whereas for passive membrane $y_p > 1$.}
\label{figTensionLess}
\end{figure}
where 
\begin{equation}
I_a(u_a) =  \int^{\infty}_{0} \frac{d\bar{t}}{\bar{t}^{3/4}} \exp{\left(-\bar{t}-\frac{u_a^2}{\bar{t}^{3/2}}\right)}.  
\end{equation}
For $u_a \rightarrow 0$, 
\begin{equation}
I_a(u_a) \simeq I_a(0) (1-b_0 |u_a|^{1/3} + b_1 |u_a|^{5/3} +b_2 |u_a|^2)   
\end{equation}
where $I_a(0)=\Gamma(1/4)$, and 
$b_0=4\,\Gamma(5/6)/\Gamma(1/4)$,  $b_1=\frac{ 2^{2/3}}{\sqrt{3\pi}} \Gamma(-\frac{5}{3})$, and $b_2= \frac{2}{\pi \, 3^{3/4}} \Gamma(-\frac{5}{12})  \Gamma(\frac{11}{12}).$
\begin{table*}
\begin{tabular}{ |p{3.8cm} p{3.2cm} p{0.9cm} p{3.2cm} p{1.0cm}|}
 \hline \hline
 \multicolumn{5}{|c|}{Correlation and Mean first-passage time }  \\
 \hline
 Membranes & \hspace{0.5cm} $\langle h^2(x,t) \rangle \sim$  & Eqs. &  $\lim_{r\to0}\langle T \rangle \sim $ & Eqs. \\
 \hline 
  Tension-dominated active  & \hspace{0.5cm} $ D_a \,\, \nu^{-1/2}\,\, t^{3/2}$ &  (\ref{ActiveHt-HtCorr})  &  $ r^{-3/4}\,h_0^{1/3}\,\, \nu^{1/12}\, D_a^{-1/6} $   & \,(\ref{TimeActiveTension}) \vspace{0.4cm}\\
 Tension-less active    & \hspace{0.5cm} $D_a \,\,  \kappa^{-1/4} \, t^{7/4}$ & \,(\ref{ActiveCorrKappa}) & $ r^{-7/8}\,h_0^{1/7}\,\, \kappa^{1/56} \,D_a^{-1/14} $ & \,(\ref{TActiveRigidity}) \vspace{0.4cm}\\
Tension-dominated passive  & \hspace{0.5cm} $D\,\nu^{-1/2}\, \,t^{1/2}$ &  \,(\ref{HtHt_ThermalNu})&  $ r^{-1/2}\,h_0^2 \, \nu^{1/2}\,\, D^{-1} $ & \,(\ref{TimePassiveTension}) \vspace{0.4cm}\\
Tension-less passive & \hspace{0.5cm} $\,D\, \kappa^{-1/4}\, t^{3/4}$ & \,(\ref{PassiveCorrKappa}) &  $r^{-3/8} \, h_0^{5/3}\,\,\kappa^{5/24}\, D^{-5/6} $ & \,(\ref{TPassiveRigidity}) \vspace{0.4cm}\\
\hline
\end{tabular}
\caption{Single point height-height correlation and mean first-passage times with active and passive noises. For tension-dominated passive membrane ($\nu\neq0$), MFPT has the largest power $h_0^2$, whereas tension-less active membrane has the lowest power $h_0^{1/7}$. We show the dependence of MFPT on the membrane properties $\nu$, $\kappa$, target height $h_0$, and resetting rate $r$ for $r \rightarrow 0.$ Our study reveals that active noise, compared with the thermal, significantly reduces the dependence on the membrane properties. }
\label{TableCorrelationMFPT}
\end{table*}

We are interested in reducing the MFPT which indeed occurs for a smaller resetting rate $r$. Thus, the MFPT is simplified as
\begin{equation}
\langle T_{\nu} \rangle_{\text{active}} \simeq \frac{b_0}{r^{3/4}} \left(\frac{h_0}{\alpha}\right)^{1/3}  \,\,\,  \text{for}  \,\,\, r \ll \left(\frac{\alpha}{h_0}\right)^{4/3}
\label{TimeActiveTension}
\end{equation}
The above scaling suggests that $\langle T_{\nu} \rangle_{\text{active}} \sim \left(a^2D_a \tau_a/\sqrt{\nu}\right)^{-1/6}.$

\subsection{\bf MFPT: Tension-less active membrane}
We next investigate the MFPT in which the membrane relaxation is dominated by $\kappa$. We first write the propagator  as 
\begin{equation}
W_{\text{active}}(h_0,t) = \frac{1}{\sqrt{\pi} \beta_t}  \exp{\left(-\frac{h_0^2}{\beta_t^2}\right)} 
\label{AMH_Propa}
\end{equation}
where $\beta_t= \beta \, t^{7/8}$ and $\beta=\sqrt{\frac{16}{21} (2^{3/4}-1)\, \Gamma(1/4) \frac{a^2D_a \tau_a }{\pi \kappa^{1/4}} }.$ We consider LT
of the above equation, which is expressed as 
\begin{equation}
\widetilde{W}_{\text{active}}(y_a,r) = \frac{1}{\sqrt{\pi} \beta} \frac{1}{r^{1/8}}\,\int^{\infty}_{0} \, \frac{d\bar{t}}{\bar{t}^{7/8}}  \exp{\left(-\bar{t}-\frac{y_a^2}{\bar{t}^{\frac{7}{4}}} \right)}
\end{equation}
where $y_a= r^{7/8}\,h_0/\beta$ and $\widetilde{W}_{\text{active}}(0,r) = \frac{1}{\sqrt{\pi} \beta} \Gamma{(1/8)}/r^{1/8}.$ Substituting the above equation in  Eq.\,(\ref{MFPT_LT_Prop}),  we obtain 
\begin{equation}
\langle T_{\kappa} \rangle_{\text{active}} = \frac{1}{r}  \left(\frac{J_a(0)}{J_a(y_a) } -1 \right)
\label{Time_Tension-LessActive}
\end{equation}
where
\begin{equation}
J_a(y_a) = \int^{\infty}_{0} \frac{d\bar{t}}{\bar{t}^{7/8}} \exp{\left(-\bar{t}-\frac{y_a^2}{\bar{t}^{7/4}}\right)}    
\end{equation}
where $J_a(0)=\Gamma(1/8)$. For $x\ll 1$, $J_a(y_a)$ can be expressed as 
\begin{equation}
J_a(y_a) \simeq J_a(0) (1-c_0 |y_a|^{1/7} + c_1 |y_a|^{9/7} -c_2 |y_a|^2) 
\end{equation}
where $c_0=1.111$,  $c_1=0.294$, and $c_2= 0.310.$ 
Keeping only the leading order, we obtain 
\begin{equation}
\langle T_{\kappa} \rangle_{\text{active}} \simeq \frac{c_0}{r^{7/8}} \left(\frac{h_0}{\beta}\right)^{1/7}   \,\,\, \text{for}  \,\,\, r \ll \left(\frac{\beta}{h_0}\right)^{8/7}
\label{TActiveRigidity}
\end{equation}

 Our study suggests that the MFPT is lowest at an intermediate value of $r$ for all cases as shown in Figs.\,(\ref{figTensionDominated}) and (\ref{figTensionLess}). As $r$ deviates from this value, the MFPT increases. Because, when $r$ is very large, the membrane moves too slowly to reach the target due to frequent resetting, and when $r$ is very small, stochastic fluctuations may cause the membrane to move far away from the target, resulting in a significantly longer MFPT.

The MFPT scales as $\langle T_{\kappa} \rangle_{\text{active}} \sim \kappa^{1/56}$ for a rigidity-dominated active membrane, whereas it scales as $\langle T_{\nu} \rangle_{\text{active}} \sim \nu^{1/12}$ for a tension-dominated active membrane, indicating that the effect of $\nu$ on MFPT is much stronger than the rigidity $\kappa$. 

 On the other hand, active noise significantly reduces the effect of $\kappa$ (or $\nu$) on MFPT compared with the thermal noise. For rigidity-dominated membrane, MFPT with active noise scales as $\langle T_{\kappa} \rangle_{\text{active}} \sim \kappa^{1/56}$ whereas with thermal noise it scales as $\langle T_{\kappa} \rangle_{\text{passive}} \sim \kappa^{5/24}$. Similarly, for tension-dominated membrane, MFPT with active noise scales as $\langle T_{\nu} \rangle_{\text{active}} \sim \nu^{1/12}$ whereas with thermal noise it scales as 
$\langle T_{\nu} \rangle_{\text{passive}} \sim \nu^{1/2}$.


\section{Discussion and Conclusions}
We study the MFPT for one-dimensional membranes under stochastic resetting with active and passive noises. Starting with the coupled equations for membrane heights and active noises, we write a Fokker-Planck equation for the joint distribution and then solve it using the method of characteristics. The explicit solution of the Fokker-Planck equation for joint distribution describes how a single height distribution depends on the single point height-height correlation ($\langle h^2 \rangle $).

Our study shows that the height-height correlation with active noise grows much faster than that with thermal noise for both the tension-dominated and tension-less membranes (Table \ref{TableCorrelationMFPT}). Across the membrane properties, the tension-less membrane shows a larger exponent of time than the tension-dominated membranes shown in Table\,\ref{TableCorrelationMFPT}.       

We derive a general relation between MFPT and a propagator with resetting (given in Eq.\,(\ref{MFPT_LT_Prop})). Using the relation and the obtained propagators for heights, we next analytically obtain the MFPT under stochastic resetting with active (or thermal) noise.  

Finally,  we demonstrate how  $\langle T\rangle$ scales with resetting rate $r$ and target height $h_0$ and how it differs between active and passive systems. Our study reveals that, starting with a very small resetting rate, $\langle T\rangle$ decreases with increasing $r$, whereas starting with a very high resetting rate, MFPT decreases with decreasing $r$. Since $\langle T \rangle$ decreases when $r$ increases to intermediate ranges from both the extrema of $r$, this indicates that there is an optimal resetting rate at which MFPT is minimized (shown in Figs.\,(\ref{figTensionLess}) and (\ref{figTensionDominated})). This is explained in the following way: In the absence of resetting, the fluctuating membrane may move so far away from the target that the first-passage time may be infinitely long. With a smaller resetting rate, the interface is reset to its initial height, which may reduce the possibility of moving the membrane far away from the target, as resetting brings it back to its initial level. With very high $r$, on the other hand, the membrane is reset so frequently that it moves too little to reach the target, causing the MFPT to grow. This indicates that at the intermediate resetting rate (the optimal resetting rate), MFPT becomes minimal.

With active dynamics, the optimal resetting rate occurs at a smaller resetting rate than that of passive systems shown in Figs.\,(\ref{figTensionLess}) and (\ref{figTensionDominated}). In terms of scaling of the target height $h_0$, a tension-dominated passive membrane has the largest value of the exponent of $h_0$ ($\sim h^2_0$), whereas a tension-less active membrane has the lowest value of exponent as  $h_0^{1/7}$ (see Table-\ref{TableCorrelationMFPT}). Our work reveals that with active noise, the scaling exponents of  $\langle T \rangle$ on $\nu$ or $\kappa$ are smaller by an order of magnitude compared with the passive system (Table-\ref{TableCorrelationMFPT}).  

This work may also serve as a benchmark for future studies that may have  real-life applications, such as the fluctuating membrane for a finite system, the nonzero resetting time, resetting with a finite velocity, etc. Within this framework, the MFPT of a fluctuating membrane in higher dimensions can be easily   generalized.

\section*{Acknowledgment}
TS would like to thank Shamik Gupta for several valuable discussions. The author thanks Mustansir Barma, Carles Blanch-Mercader, and Pierre Sens for useful discussions.
TS also acknowledges the support provided by a grant from ITMO Cancer, PSCI.
\section*{Appendix} 

\section*{Appendix-A: Derivation of the propagator}

\subsection*{Fokker-Planck equation}
 We consider $\mathbf{p}^{T}=\{p_i\}$ and $\mathbf{q}^{T} =\{q_i\}$, which are the Fourier variables corresponding to $\tilde{h}=\{h_i\}$ and $\tilde{v}=\{v_i\}$, respectively. Let us write the Fourier transform in the following way
\begin{equation}
\widehat{\mathbf{W}}(\mathbf{p}^{T},\mathbf{q}^{T},t|\psi^{0}) = \int d\tilde{h} \int d\tilde{v} \,\,  e^{-i(\mathbf{p}^{T} \tilde{h}+\mathbf{q}^{T}\tilde{v})} \mathbf{W} (\tilde{h},\tilde{v},t|\psi^{0})
\end{equation}
where $(\mathbf{p}^{T})_{1\times L}$, $(\mathbf{q}^{T})_{1\times L}$, and $(h)_{L\times 1}$ and $(v)_{L\times 1}$ matrices. Taking into account the initial conditions, we obtain $
\widehat{\mathbf{W}}_{0}(\mathbf{p^0}^{T},\mathbf{q^0}^{T},0|\psi^{0}) = \int d\tilde{h} \int d\tilde{v} \,\,  e^{-i(\mathbf{p^{0}}^{T} \tilde{h}+\mathbf{q^{0}}^{T}\tilde{v})} \,\, \delta(\tilde{h}-\tilde{h}^{0}) \, \delta(\tilde{v}-\tilde{v}^{0})  = e^{-i\,\mathbf{p^{0}}^{T} \tilde{h}^{0}-i\,\mathbf{q^{0}}^{T}\tilde{v}^{0}}.$ We now obtain a Fokker-Planck equation in Fourier space as
\begin{eqnarray}
\frac{\partial \mathbf{\widehat{W}}}{\partial t} &&= -\sum_{i,j} \Lambda_{ij}\, p_i \,\, \frac{\partial \mathbf{\widehat{W}}}{\partial p_j} + a \sum_{ij} p_i \frac{\partial \mathbf{\widehat{W}}}{\partial q_i} \nonumber \\ && - \sum_{ij} \Lambda_{ij}' q_i \, \frac{\partial \mathbf{\widehat{W}}}{\partial q_i} - \sum_{ij} D_{ij} p_i p_j \mathbf{\widehat{W}} -  \sum_{ij} D^{a}_{ij} q_i q_j \mathbf{\widehat{W}}. \nonumber      
\end{eqnarray}

\subsection{Method of characteristics}

We employ the method of characteristics and get \begin{equation}
\frac{d \mathbf{\widehat{W}}}{d t} = \frac{\partial \mathbf{\widehat{W}}}{\partial t} + \sum_{i} \left(\frac{\partial \mathbf{\widehat{W}}}{\partial p_i}  \frac{dp_i}{dt} + \frac{\partial \mathbf{\widehat{W}}}{\partial q_i}  \frac{dq_i}{dt}\right).
\end{equation}
Along the characteristic line, we have 
\begin{eqnarray}
&& \frac{d p_i}{d t} = \sum_{j} \Lambda_{ij} p_j,\\   
&& \frac{d q_i}{d t} = \sum_{j} \Lambda'_{ij} q_j - a \sum_{ij} p_i. \end{eqnarray}
Solving the above equations, we obtain  
\begin{equation}
\mathbf{p}(t) = e^{t \Lambda} \, \mathbf{p^{0}}   
\label{p_time}
\end{equation}
and
\begin{equation}
\mathbf{q}(t) = e^{t \Lambda'} \, \mathbf{q^{0}} +a \left(\frac{e^{\Lambda t}-e^{\Lambda' t}}{\Lambda'-\Lambda}\right) \mathbf{p}^{0}.  
\label{Sol_q}
\end{equation}
In the above equations, $\mathbf{p}(t)$ and $\mathbf{q}(t)$ evolve along the characteristic line. The total time derivative of FPE is obtained as 
\begin{eqnarray}
&&\frac{d \mathbf{\widehat{W}}}{dt} = -\frac{1}{2} \sum_{ij} 2 D_{ij} \,\, p_i \, p_j \mathbf{\widehat{W}} -\frac{1}{2} \sum_{ij} 2 D^{a}_{ij}\, q_i \,q_j \mathbf{\widehat{W}} \\
&& \frac{d (\log{\mathbf{\widehat{W}}})}{dt} = -\frac{1}{2} \left( \mathbf{p}^{T}2D\, \mathbf{p} + \mathbf{q}^{T} 2D_{a}\, \mathbf{p}\right).
\label{Eq_Charac}
\end{eqnarray} 
Let us calculate the argument in the above equation as
$\mathbf{p}^{T}2D\mathbf{p} + \mathbf{q}^{T} 2D_{a} \mathbf{p}$ in terms of $\mathbf{p}^{0}$ and $\mathbf{q}^{0}$. Using $\mathbf{q}$ from Eq.(\ref{Sol_q}), we obtain  
\begin{eqnarray}
\mathbf{q}^{T} D_{a} \mathbf{q} &&= (\mathbf{p^{0}})^{T} Q_1\mathbf{p}^{0} +  (\mathbf{q^{0}})^{T} Q_2\mathbf{q}^{0} + 2 (\mathbf{p^{0}})^{T} Q_3 \mathbf{q}^{0} \nonumber \\
\label{Cha_q_Matrix}
\end{eqnarray}
where 
\begin{equation}
(Q_1)_{L\times L} = a^2 \left[2D_{a}\,\left(\frac{e^{\Lambda t}-e^{\Lambda' t}}{\Lambda'-\Lambda} \right)^2\right]_{L\times L},    
\end{equation}

\begin{equation}
(Q_2)_{L\times L} = (2D_a\, e^{2\Lambda' t})_{L\times L},
\end{equation}
and 
\begin{equation}
(Q_3)_{L \times L} = a \left[2D_a \left(\frac{e^{\Lambda t}-e^{\Lambda' t}}{\Lambda'-\Lambda} \right) e^{\Lambda' t}\right]_{L \times L}.
\end{equation}
Similarly Eq.\,(\ref{p_time}) leads to
\begin{equation}
\mathbf{p}^{T} 2D\mathbf{p} = \mathbf{p^{0}}^{T} (e^{\Lambda t} 2D e^{\Lambda t}) \,\, \mathbf{p^{0}} 
\label{Cha_p_Matrix}
\end{equation}
Adding Eqs.\,(\ref{Cha_q_Matrix}) and (\ref{Cha_p_Matrix}), we obtain 
\begin{eqnarray}
\mathbf{p}^{T}2D\mathbf{p} + \mathbf{q}^{T} 2D_{a} \mathbf{p} = (\mathbf{p^{0}})^{T} && (Q_1+Q'_1)\, \mathbf{p}^{0} +  (\mathbf{q^{0}})^{T} Q_2\mathbf{q}^{0} \nonumber \\ && + 2 (\mathbf{p^{0}})^{T} Q_3 \, \mathbf{q}^{0}.
\label{Argu}
\end{eqnarray}
We now substitute Eq.\,(\ref{Argu}) in Eq.\,(\ref{Eq_Charac}), and obtain
\begin{eqnarray}
\log \left(\frac{\mathbf{\widehat{W}}}{\mathbf{\widehat{W}_0}}\right) &&=-\frac{1}{2} \left[ (\mathbf{p^{0}})^{T} R_1\,\mathbf{p}^{0} +  (\mathbf{q^{0}})^{T} R_2\mathbf{q}^{0} + 2 (\mathbf{p^{0}})^{T} R_3 \, \mathbf{q}^{0} \right] \nonumber \\
\label{Sol_Charac1}
\end{eqnarray}
where $\mathbf{\widehat{W}_0}$ is the Fourier transform of the initial height configuration and velocity profiles. In Eq.(\ref{Sol_Charac1}), we have 
\begin{eqnarray}
&& R_1 = \int^{t}_{0} dt' \, (Q_1(t')+Q'_1(t')), \\
&& R_2 = \int^{t}_{0} dt' Q_2(t'), \\
&& R_3 = \int^{t}_{0} dt' Q_3(t').
\label{R1R2R3}
\end{eqnarray} Using $\mathbf{\widehat{W}_0}$ (as shown above) and Eq.(\ref{Sol_Charac1}), we obtain 
\begin{eqnarray}
\widehat{\mathbf{W}} && = \exp{(-\frac{1}{2}   [ \mathbf{p^0}^T R_1\mathbf{p}^0   +  \mathbf{q^0}^T R_2\mathbf{q}^0 + 2 \mathbf{p^0}^T R_3 \mathbf{q}^0]) }  \nonumber \\ && \times \exp{(-i(\mathbf{p^0}^T  \tilde{h}^{0}  + \mathbf{q^0}^T\tilde{v}^0)))}. \nonumber \\
\end{eqnarray}
Below, we express $\mathbf{p^0}^{T}$, $\mathbf{q^0}^{T}$, $\mathbf{p^0}$, and $\mathbf{q^0}$ in terms of $\mathbf{p}^{T}$ and $\mathbf{q}^{T}$. Therefore, we write the arguments in the above equations as
$ \mathbf{p^{0}}^T R_1 \mathbf{p^{0}} = \mathbf{p}^T (e^{-\Lambda t} R_1 e^{-\Lambda t})\, \mathbf{p} $, and 
\begin{eqnarray}
\mathbf{q^{0}}^T R_2 \mathbf{q^{0}} 
&&= \mathbf{q}^{T}(e^{-2\Lambda' t} R_2)\mathbf{q} +  \mathbf{p}^{T} a^2 e^{-2(\Lambda'+\Lambda) t} R_2  f^2  \mathbf{p} \nonumber \\ && - 2 \mathbf{p}^{T} \left(a e^{-2\Lambda' t} R_2 f e^{-\Lambda t}\right) \mathbf{q}. \label{qTq}
\end{eqnarray} The other term can be obtained as
\begin{eqnarray}
2 \mathbf{p^0}^{T} R_3 \mathbf{q^0} && = 2 \mathbf{p}^{T} ( e^{-(\Lambda+\Lambda')t}R_3) \mathbf{q} \nonumber  \\ + 2\mathbf{p}^{T}  && (-a e^{-(2\Lambda+\Lambda') t}R_3 f) \mathbf{p}. \label{pTq}
\end{eqnarray}
Combining the above terms, we get
\begin{eqnarray}
&&\exp{\left[-\frac{1}{2}\left(\mathbf{p^0}^T R_1 \mathbf{p^0} + \mathbf{q^0}^T R_2 \mathbf{q^0} +2 \mathbf{p^0}^T R_3 \mathbf{q^0}\right)\right]}  \nonumber \\ &&=  \exp{\left[-\frac{1}{2}(\mathbf{p}^T\,\mathbf{M_1} \mathbf{p} + \mathbf{q}^T \mathbf{M_2} \mathbf{q} + 2\, \mathbf{p}^T \mathbf{M_3} \mathbf{q})\right]}.  
\end{eqnarray}
The contribution to $\mathbf{p}^{T} \mathbf{p}$ also comes from $\mathbf{q^0}^{T}\mathbf{q^0}$ and $\mathbf{p^0}^{T}\mathbf{p^0}.$ Therefore, we have three terms in $\mathbf{M}_1.$
\begin{eqnarray}
\mathbf{M}_1 && = (e^{-\Lambda t} R_1 e^{-\Lambda t} + a^2\, e^{-2(\Lambda+\Lambda')t}\, f^2(t) R_2 \nonumber \\ && - 2\, a\, e^{-(2\Lambda+\Lambda')t} f(t) \, R_3)_{L\times L} .
\label{M1}
\end{eqnarray}
Let us now evaluate the integrations for $R_1$, $R_2$ and $R_3$ which are given in Eq.\,(\ref{R1R2R3}). Thus, we now have 
\begin{eqnarray}
R_1 =(2D+\frac{2a^2 D_a}{(\Lambda'-\Lambda)^2}&& )(\frac{e^{2\Lambda t}-1}{2\Lambda} )  + \frac{2a^2 D_a}{(\Lambda'-\Lambda)^2} (\frac{e^{2\Lambda' t}-1} {2\Lambda'}) \nonumber \\ && -4 \frac{a^2 D_a}{(\Lambda'-\Lambda)^2}\frac{(e^{(\Lambda+\Lambda')t}-1)}{\Lambda'+\Lambda}, \end{eqnarray}
\begin{eqnarray}
R_2 &&= 2D_a \, \left(\frac{e^{2\Lambda't}-1}{2\Lambda'}\right)
\end{eqnarray}
and
\begin{eqnarray}
R_3 &&= a\frac{2D_a}{\Lambda'-\Lambda} \left( \frac{e^{(\Lambda+\Lambda')t}-1}{(\Lambda+\Lambda')} - \frac{e^{2\Lambda't}-1}{2\Lambda'}\right)
\end{eqnarray}
Substituting the expressions for $R_1$, $R_2$ and $R_3$ in Eq.\,(\ref{M1}),  we obtain 
\begin{eqnarray}
\mathbf{M}_1 \, \delta_{ij} = 2D_{\text{tot}} \left(\frac{1-e^{-2\Lambda t}}{2\Lambda} \right) + && \frac{a^2 2 D_a}{(\Lambda'-\Lambda)^2} \{\frac{2\,(e^{-(\Lambda+\Lambda')t}-1)}{(\Lambda'+\Lambda)}\, \nonumber \\ && + \frac{\left(1-e^{-2\Lambda' t} \right)}{2\Lambda'} \}   
\end{eqnarray}
where $D_{\text{tot}}=D + \frac{a^2D_a}{(\Lambda'-\Lambda)^2}$. Similarly, we obtain
\begin{equation}
\mathbf{M}_2 \, \delta_{ij} = 2D_a \left(\frac{1-e^{-2\Lambda't}}{2\Lambda'}\right)
\end{equation}
and
\begin{equation}
\mathbf{M}_3 \, \delta_{ij} = a\frac{2D_a}{(\Lambda'-\Lambda)} \left(\frac{1-e^{-(\Lambda+\Lambda')t}}{\Lambda+\Lambda'}-\frac{1-e^{-2\Lambda' t}}{2\Lambda'}\right).
\end{equation}
\subsection{Propagator}
Let us write  $\widehat{\mathbf{W}}[\mathbf{p}, \mathbf{q},t|\psi^{0}]$ in terms of $\mathbf{M}_1$, $\mathbf{M}_2$ and $\mathbf{M}_3$ as
\begin{widetext}
\begin{eqnarray}
\mathbf{\widehat{W}}[\mathbf{p},\mathbf{q},t|\psi_{0}] = \exp{[-i\,(\mathbf{p}^T e^{-\Lambda t} \tilde{h}^{0} + \mathbf{q}^T e^{-\Lambda' t} \tilde{v}^{0})]}&& \exp{(-i\,( -a\,\mathbf{p}^T f(t)\,  e^{-(\Lambda'+\Lambda)t} \tilde{v}^{0})}) \nonumber \\  &&  \times \exp{(-\frac{1}{2} (\mathbf{p}^T\,\mathbf{M_1} \mathbf{p} + \mathbf{q}^T \mathbf{M_2} \mathbf{q}  + 2\, \mathbf{p}^T \mathbf{M_3} \mathbf{q}))} 
\end{eqnarray}
We now consider the inverse Fourier transform and integrate over $p$ and $q$. Thus, we obtain 
\begin{eqnarray}
&& \mathbf{W}[\tilde{h},\tilde{v},t|\psi_{0}] = \int \frac{d \mathbf{p}}{(2\pi)^{L}} \int \frac{d \mathbf{q}}{(2\pi)^{L}} \,\, e^{i(\mathbf{p}^{T}\tilde{h}+\mathbf{q}^{T}\tilde{v})} \,\, \mathbf{\widehat{W}}[\mathbf{p},\mathbf{q},t|\psi^{0}]\nonumber \\
 && =  \frac{1/(2\pi)^L}{\sqrt{\det{(\mathbf{M}_1\mathbf{M}_2)}}} \exp{\left[-\frac{1}{2}\left(\Delta\tilde{h}\,\mathbf{M}^{-1}_1 \Delta\tilde{h}\right)-\frac{1}{2}\left(\Delta\tilde{v}^{T}\mathbf{M}^{-1}_2\Delta\tilde{v}\right)-\frac{1}{2}\Delta\tilde{h}^{T} \left(\frac{\mathbf{M}_2}{\mathbf{M}_3^2}\right)^{-1} \Delta\tilde{h} + \Delta\tilde{h}^{T}\left(\frac{\mathbf{M}_2}{\mathbf{M}_3}\right)^{-1} \Delta\tilde{v}\right]}  \nonumber\\
 \end{eqnarray}
\end{widetext}
where $\Delta\tilde{h}=\tilde{h}-e^{-\Lambda t}\,h^0$, and $\Delta\tilde{v}=\tilde{v}-e^{-\Lambda' t}\,v^0$. Integrating over $\tilde{v}$, we obtain the marginal distribution
\begin{eqnarray}
\mathbf{W}(\tilde{h},t|h_0) &&=\int d \tilde{v} \int dv_0\,\, \mathbf{W}(\tilde{h}, \tilde{v},t|\psi_{0}) \nonumber \\
&& = \frac{\exp{[-\frac{1}{2}(\tilde{h}-e^{-\Lambda t} h^0)^T \mathbf{M}^{-1}_1 (\tilde{h}-e^{-\Lambda t}h^0)]}}{(2\pi)^{L/2} \sqrt{\det{\mathbf{M}}_1}}.\nonumber \\
\label{ProbDist_Ht1}
\end{eqnarray}

\subsection{Height-height correlation in terms of operator} 
Let us start with the non-stationary  active noise. Setting $t_0=0$, we obtain
\begin{eqnarray}
\langle v_{i}(t)v_{j}(t') \rangle &&= e^{-\Lambda'(t+t')} \int^{t}_{t_0} ds \int^{t}_{t_0} ds' \, e^{\Lambda'(s+s')} \langle \Gamma_{i}(s) \Gamma_{j}(s') \rangle \nonumber \\
&& =\delta_{ij} \, D_a \, e^{-\Lambda'(t+t')} \left( \frac{e^{2\Lambda't}-1}{\Lambda'}\right) \,\,\, \text{when\,\,} \, t'>t. \nonumber \\
\end{eqnarray}
Taking into account the other part, i.e., $t>t'$, we obtain
\begin{eqnarray}
\langle v_{i}(t)v_{j}(t') \rangle &&= \frac{D_a}{\Lambda'} \left(e^{-\Lambda'|t-t'|}-e^{-\Lambda'(t+t')} \right)
\label{non-ActiveNoise}
\end{eqnarray}
and 
\begin{equation}
\langle \eta_{i}(t)\eta_{j}(t') \rangle = 2D \delta_{ij} \delta(t-t').   \end{equation}
We first study the height-height correlation when the velocity-velocity correlation is non-stationary (i.e., $t_0 =0$). We obtain\begin{eqnarray}
 && \langle h_{i}(t) h_{j}(t) \rangle = e^{-2\Lambda t} (\int^{t}_{0} dt' e^{\Lambda t'} \int^{t}_{0} dt'' e^{\Lambda t''} \langle \eta_{i}(t') \eta_{j}(t'') \rangle \nonumber \\
 && +a^2\, \int^{t}_{0}dt' e^{\Lambda t'} \int^{t}_{0}dt'' e^{\Lambda t''} \langle v_{i}(t') v_{j}(t'') \rangle  ) \nonumber\\ 
 &&= \delta_{ij}\,D_{\text{tot}} \left(\frac{1-e^{-2\Lambda t}}{\Lambda} \right) + \delta_{ij}\,\frac{2\,a^2 D_a}{(\Lambda'-\Lambda)^2} \{\frac{2\,(e^{-(\Lambda+\Lambda')t}-1)}{(\Lambda'+\Lambda)}\, \nonumber \\
 && + \frac{\left(1-e^{-2\Lambda' t} \right)}{2\Lambda'}  \}  \\ 
 && = \sum_{ij} (M_{1})_{ij} \, \delta_{ij}.
 \end{eqnarray}
The covariance matrix from the Fokker-Planck equation and the height-height correlation from the coupled Langevin equations are found to be the same. The relation in the above equation is obtained with the non-stationary active noise, which also holds for the stationary-state active noise.

\section*{Appendix-B}


\subsection{MFPT: Tension-dominated passive membrane}
The height-height correlation can be written as
\begin{equation}
\langle h(x,t)^2 \rangle_{\text{passive}} =  \int \frac{dq}{(2\pi)} \int \frac{dq'}{(2\pi)} \,\, e^{i (q+q') x}\, \langle h(q,t) h(q',t)\rangle . 
\end{equation}
In the above equation, we consider the membrane fluctuations driven by spatially and temporally uncorrelated thermal noise. The Fourier transform of the correlation can be expressed as $\langle\eta(q,t) \eta(q',t') \rangle = 2 D (2 \pi) \delta(q+q') \delta(t-t')$. Using this, we obtain
\begin{eqnarray}
\langle h(x,t)^2 \rangle_{\text{passive}} = 2D \,  \int^{\infty}_{-\infty} \frac{dq}{(2\pi)} \,\, \int^{t}_0 dt' \,\,   e^{-2\nu t' q^2  } .
\end{eqnarray}
We carry out the integration over $q$ and $t'$ in the above equation and obtain 
\begin{eqnarray}
\langle h(x,t)^2 \rangle_{\text{passive}}  = D \sqrt{\frac{2t}{\pi \nu}}. \end{eqnarray} Substituting the above equation in Eq.\,(\ref{EqPropagator}), we obtain 
\begin{equation}
W_{\text{passive}}(h_0,t) = \frac{1}{\sqrt{\pi \alpha^2_p(t)}} \, e^{-\frac{h_0^2}{ \alpha^2_p(t)}\, }   
\end{equation}
where $\alpha^2_T(t) =2D \sqrt{\frac{2t}{\pi \nu}}.$ We consider the LT of the propagator for passive membrane dynamics with thermal noise. The LT of the above equation can be obtained as
\begin{equation}
\widetilde{W}_{\text{passive}}(u_p, r) = \frac{1}{\sqrt{\pi\alpha^2_p}} \frac{1}{r^{3/4}} \int^{\infty}_{0} \frac{d\bar{t}}{\bar{t}^{1/4}} \exp{\left(-\bar{t}-\frac{u_p^2}{\sqrt{\bar{t}}}\right)}  \end{equation}
where $\bar{t}= r t$, $\alpha_T^2 = 2D \sqrt{2/(\pi \nu)}$ and $u_p=r^{1/4}h_0/\alpha_T$. Therefore, our MFPT can be written as
\begin{equation}
\langle T_{\nu} \rangle_{\text{passive}} = \frac{1}{r}  \left(\frac{I_p(0) }{I_p(u_p)} -1 \right)
\label{Time_TDPM}
\end{equation}
where
\begin{equation}
I_p(u_p)=\int^{\infty}_{0} \frac{d\bar{t}}{\bar{t}^{1/4}} \exp{\left(-\bar{t}-\frac{u_p^2}{\sqrt{\bar{t}}}\right)}. 
\end{equation}
We carry out the integration for $u_p=0$ and obtain  $I_p(0)=\Gamma(\frac{3}{4}).$
From the above equation, we obtain
 \begin{equation} 
 I_p(u_p) \simeq I_p(0) \left(1-|u_p|^2\frac{\Gamma(\frac{1}{4})}{\Gamma(\frac{3}{4})} \right)
 \end{equation}
for  $u_p \ll 1. $
A simpler form for the MFPT can be expressed as
\begin{equation}
\langle T_{\nu} \rangle_{\text{passive}} \simeq  \frac{\Gamma(\frac{1}{4})}{\Gamma(\frac{3}{4})} \frac{1}{r^{1/2}} \,\,\,  \left(\frac{h_0}{\alpha_T}\right)^2  \,\,\,  \text{for}  \,\,\, r \ll \left(\alpha_T/h_0\right)^{4}.
\label{TimePassiveTension}
\end{equation}
The expression of the MFPT $\langle T_{\nu} \rangle_{\text{passive}} \rightarrow \infty $ when $u_p \rightarrow \infty$. The MFPT of a passive membrane scales as $\langle T_{\nu} \rangle_{\text{passive}} \sim \nu^{1/2}.$


\subsection{MFPT: Tension-less passive membrane}
Similarly, with the thermal noise, we derive the height-height correlation for the rigidity-dominated membrane as
\begin{eqnarray}
\langle h(x,t)^2 \rangle_{\text{passive}} \, &&= \frac{D}{3\pi^{3/4}}\, \left[ \frac{(2t)^3}{\pi \kappa} \right]^{1/4}
\end{eqnarray}
where we consider $\nu=0.$ Substituting the above equation in equation  (\ref{EqPropagator}), we write the propagator as 
\begin{eqnarray}
W_{\text{passive}}(h_0,t) = \frac{1}{\sqrt{\pi}} \frac{1}{\beta_T(t)} \exp{\left(-\frac{h_0^2}{\beta_T(t)^2}\right)}
\end{eqnarray}
where $\beta_T(t)=\beta_T\,\, t^{3/8}$ 
with $\beta_T= \sqrt{\frac{2D}{3\pi^{3/4}}}\, \left[ \frac{8}{\pi \kappa} \right]^{1/8}.$ 
The Laplace transform of the above propagator can be written as 
\begin{eqnarray}
\widetilde{W}_{\text{passive}} (y_p, r) =  \frac{1}{\sqrt{\pi}} \frac{1}{\beta_T \, r^{5/8}} J_p(y_p) \end{eqnarray}
where $y_p= r^{3/8} h_0/\beta_T$, and 
\begin{equation}
J_p(y_p) =\int^{\infty}_0 \frac{d\bar{t}}{\bar{t}^{3/8}}\, e^{-\bar{t}}\,  \exp{\left(-\frac{y_p^2}{\bar{t}^{3/4}}\right)} .   
\end{equation}
Using the definition of the Gamma function, we obtain 
$ \widetilde{W}_{\text{passive}} (0, r) =  \frac{1}{\sqrt{\pi}} \Gamma\left(\frac{5}{8}\right)/(\beta_T \, r^{5/8})$ and  
\begin{equation}
\langle T_{\kappa} \rangle_{\text{passive}} = \frac{1}{r}  \left(\frac{J_p(0) }{J_p(y_p)} -1 \right).
\label{Time_TLPM}
\end{equation} 
Considering $ y_p \ll 1$, we obtain $ \langle T_{\kappa} \rangle \simeq \frac{1}{r} \,\, \left[(1-c_0 |y_p|^{5/3} + c_1 |y_p|^2)^{-1}-1\right]$  where $c_0 = 4.656$, and $c_1=6.077.$
For $r \ll (\beta_T/h_0)^{8/3}$ , we obtain
\begin{equation}
\langle T_{\kappa} \rangle_{\text{passive}} \simeq \frac{c_0}{r^{3/8}} \,\, \left( \frac{h_0}{\beta_T} \right)^{5/3} .
\label{TPassiveRigidity}
\end{equation}
The above equation suggests that for the passive system, the MFPT strongly depends on the rigidity $\kappa$ as $\langle T_{\kappa} \rangle_{\text{passive}} \sim \kappa^{5/24}.$


\end{document}